\documentclass[twocolumn,showpacs,preprintnumbers,amsmath,amssymb,prb,aps]{revtex4-2}
\usepackage{esvect}
\usepackage[usenames, dvipsnames]{color}
\usepackage{comment}
\usepackage{epsfig}% Include figure files
\usepackage{graphicx}% Include figure files
 \usepackage{float} % for pinning a figure
\usepackage{gensymb}% for degree
\usepackage[colorlinks=true,linkcolor=red,citecolor=blue,urlcolor  = blue]{hyperref}
\usepackage{dcolumn}   % needed for some tables
\usepackage{bm}        % for math
\usepackage{ulem}
\begin{document}

%Title of paper
\title{Evidence of electron correlation and unusual spectral evolution in an exotic superconductor, PdTe}

\author{Ram Prakash Pandeya,$^1$ Arindam Pramanik,$^1$ Pramita Mishra,$^1$ Indranil Sarkar,$^2$ A. Thamizhavel$^1$ and Kalobaran Maiti$^1$}

\altaffiliation{Corresponding author: kbmaiti@tifr.res.in}

\affiliation{$^1$ Department of Condensed Matter Physics and Materials Science, Tata Institute of Fundamental Research, Homi Bhabha Road, Colaba, Mumbai - 400 005, INDIA \\
$^2$ Deutsches Elektronen-Synchrotron DESY, Notkestrasse 85, D - 22607 Hamburg, Germany
}

\date{\today}

\begin{abstract}
We study the electronic structure of an exotic superconductor, PdTe employing depth-resolved high resolution photoemission spectroscopy and density functional theory. The valence band spectra exhibit large density of states at the Fermi level with flat intensity in a wide energy range indicating highly metallic ground state. The Pd 4$d$-Te 5$p$ hybridization is found to be strong leading to a highly covalent character of the itinerant states. Core level spectra exhibit several features including the signature of plasmon excitations. Although the radial extension of the 4$d$ orbitals is larger than 3$d$ ones, the Pd core level spectra exhibit distinct satellites indicating importance of electron correlation in the electronic structure which may be a reason for unconventional superconductivity observed in this system. The depth-resolved data reveal surface peaks at higher binding energies in both, Te and Pd core level spectra. Interestingly, core level shift in Te-case is significantly large compared to Pd although Te is relatively more electronegative. Detailed analysis rules out applicability of the charge transfer and/or band-narrowing models to capture this scenario. This unusual scenario is attributed to the reconstruction and/or vacancies at the surface. These results reveal the importance of electron correlation and surface topology for the physics of this material exhibiting Dirac fermions and complex superconductivity.
\end{abstract}

%\pacs{82.80.Pv, 73.20.At, 74.25.Jb}

\maketitle

\section{Introduction}

Topological materials exhibiting superconductivity are emerging as the new frontier of research in the condensed matter physics. Despite immense interest in this field, material candidates found in this category are barely few \cite{TISC-1,TISC-2,TISC-3,TISC-4}. In most of the cases, the Dirac fermions appear at high binding energy and do not take part in superconductivity thereby, making it highly challenging to identify topological superconductivity experimentally. A recent study suggested that PdTe is a candidate material for unconventional superconductivity as well as non-trivial topology \cite{PdTe-SREP-Chapai}. Another study showed signature of bulk nodal and surface nodeless Cooper pairing in PdTe \cite{PdTe-PRL-Hasan}. Clearly, detailed study of the surface and bulk properties in such systems are crucial to understand the physics of the exoticities of these materials. Due to significant covalent character of the bonds in most such materials, breaking of translational symmetry at the surface leads to a lower dimensional atomic arrangement with broken and/or unsaturated bonds at the surface. Such atomic arrangement reduces its energy by re-bonding with the atoms available close to the surface or by rearranging the atomic positions of the surface atoms which generally leads to different electronic states at the surface layer \cite{Surface-Reconstruction}. Understanding such aspects of surface-bulk differences in the electronic structure is, in general, fascinating from both basic physics \cite{TI-Surface} and practical applications \cite{Surface-Applications}. Among varied surface sensitive techniques utilized for studying the nature of the surfaces \cite{Surface}, the photoelectron emission spectroscopy has played a vital role due to tunability of the technique for a wide range of depth from the exposed surface (depth-resolved) with very high energy resolution \cite{Surface-PES, Hufner}.

\begin{figure}
\includegraphics[width=\linewidth]{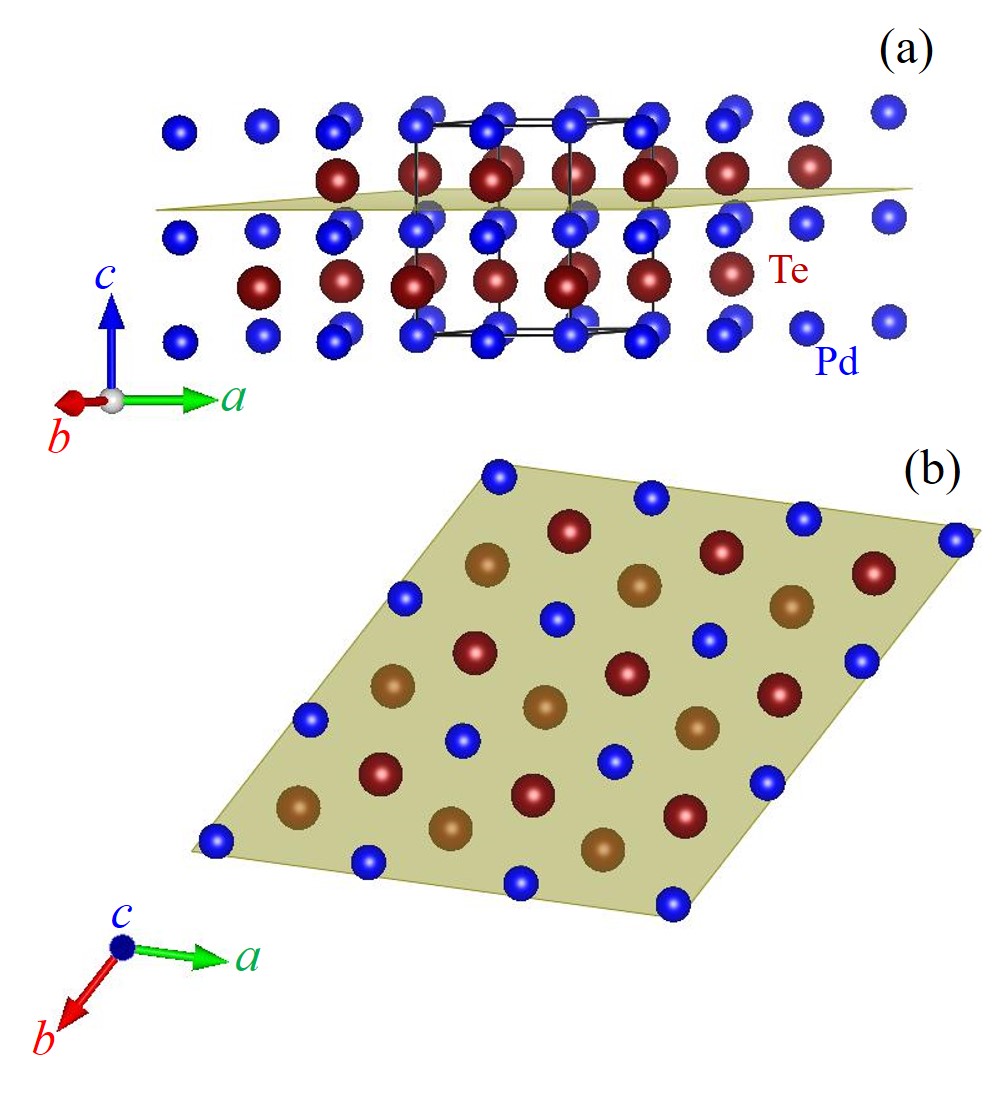}
\vspace{-4ex}
\caption{Crystal structure of PdTe exhibiting (a) ABAC-stacking along $c$-direction and (b) $ab$-plane showing hexagonal structure in each layer. The shaded area is a $ab$-plane.}
\label{str}
\end{figure}

In this paper, we report our results of investigation of the electronic structure of PdTe using depth-resolved photoemission spectroscopy and density functional theory (DFT). PdTe forms in hexagonal structure with $P6_3/mmc$ space group as shown in Fig. 1. In this structure, Pd and Te atoms are arranged in ABAC-stacking configuration as shown in Fig. \ref{str}(a). The structure in $ab$-plane is hexagonal as shown in Fig. \ref{str}(b). PdTe show superconducting phase transition at $\sim$ 4.5 K, which was believed to be of conventional (BCS)-type \cite{Brajesh-PdTe, Karki-PdTe, Karki-FePdTe, Jha-NiPdTe, PdTe-Ram}. Although, the superconductivity in PdTe was discovered a long time back \cite{Matthias-PdTe}, the scientific interest in this compound surged due to crucial similarities between this compound and unconventional superconductivity in Fe-based chalcogenide and pnicotgen compounds \cite{HHosono2015, Ekuma-Wannier-PdTe}. Using experimentally measured physical property results and theoretically calculated electron-phonon coupling constant, it has been found that the conventional electron-phonon coupling mechanism with intermediate coupling strength can explain the superconductivity in this compound \cite{Cao-DFT-PdTe}. There are many theoretical electronic structure studies available in the literature for this compound \cite{Ekuma-Wannier-PdTe, Cao-DFT-PdTe}. Most recently, the finding of Dirac bands and signature of unconventional superconductivity led to resurgence of interests in this material \cite{PdTe-PRL-Hasan}. We have carried out a detailed electronic structure study of this compound using depth-resolved photoemission spectroscopy, and find interesting surface-bulk differences and evidence of finite electron correlation in the electronic structure.

\section{Experimental Method}

We prepared high quality crystals of PdTe using modified Bridgman technique. Sample purity and crystallinity were confirmed using energy dispersive analysis of $x$-rays, powder $x$-ray diffraction and Laue $x$-ray diffraction techniques. No trace of impurity were observed \cite{PdTe-Ram}.  A sharp diamagnetic response was observed below 4.5 K exhibiting evidence of superconducting phase \cite{Brajesh-PdTe,Karki-PdTe}.

Photoemission spectroscopy (PES) measurements were carried out using monochromatic laboratory sources [He {\scriptsize I} (h$\nu$ = 21.2 eV), He {\scriptsize II} (h$\nu$ = 40.8 eV) and Al K$\alpha$ ($h\nu$ = 1486.6 eV)] and R4000 electron detection system. Both ultra-violet PES (UVPES) and conventional $x$-ray PES (CXPS) measurements were performed on the (0001) surface cleaved in ultrahigh vacuum. It is to note here that although PdTe forms in layered structure (see Fig. \ref{str}), our samples were hard to cleave. We used top-post removal method for cleaving using a specially designed sample mounting. The cleaved surface was found clean. The chamber pressure during the measurements was maintained at 1$\times$10$^{-10}$ Torr. The hard $x$-ray PES (HAXPES) measurements were carried out at P09 beamline of Petra-III, DESY, Germany ($h\nu$ = 5947 eV) on cleaved surface at 2$\times$10$^{-10}$ Torr pressure. Energy resolutions for UVPES, CXPS and HAXPES measurements were 5 meV, 400 meV and 150 meV, respectively. While the CXPS measurements were carried out at both normal emission (NE) and 60$^{\circ}$ angled emission geometries to change the surface sensitivity, the HAXPES measurements were performed in NE geometry only for the collection of the most bulk sensitive spectra.

The electronic band structure (DFT) calculations were carried out using full potential linearized augmented plane wave (FLAPW) method as implemented in Wien2k software \cite{Wien2k}. The convergence to the ground state was achieved by fixing the energy convergence criterion to 10$^{-4}$ Rydberg using 1000 $k$-points in the Brillouin zone. We have used the Perdew-Burke-Ernzerhof generalized gradient potential approximation (PBE-GGA) exchange potential for the DFT calculation \cite{PBE_GGA}. The structural parameters were taken from the simulation of room temperature powder $x$-ray diffraction pattern \cite{PdTe-Ram}.

%%%%%%%%%%%%%%%%%%%%%%%%%%%%%%%%%%%%%%%%%%%

\section{Results and Discussions}

\begin{figure}
\includegraphics[width=\linewidth]{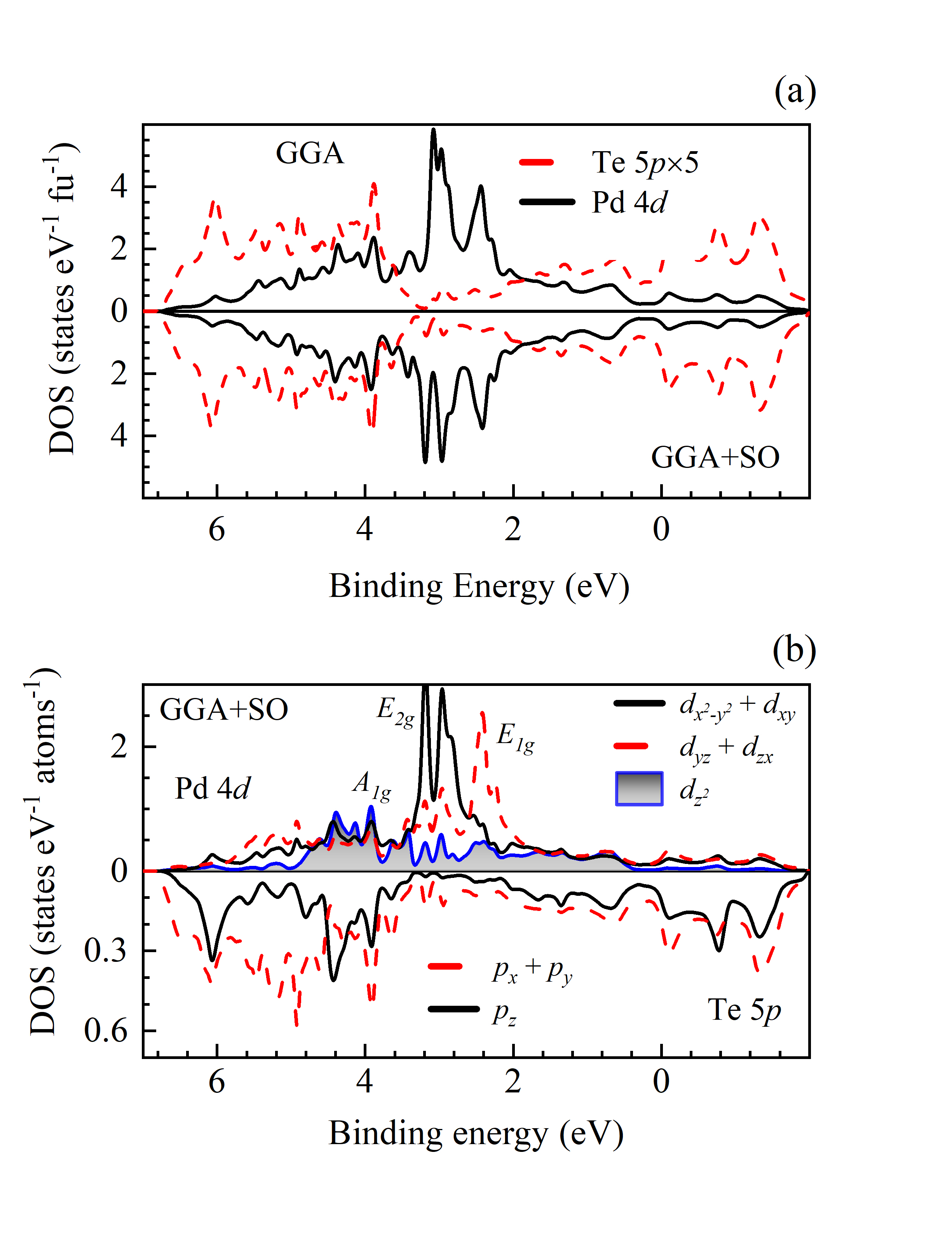}
\vspace{-4ex}
\caption{(a) Te 5$p$ (dashed line) and Pd 4$d$ (solid line) PDOS using GGA (upper panel) and GGA+SO (lower panel) method. Te 5$p$ PDOS is multiplied by 5 for better comparison. (b) Upper panel: Pd 4$d_{x^2-y^2}+d_{xy}$ (solid line), 4$d_{yz}$+4$d_{xz}$ (dashed line) and 4$d_{z^2}$ (area). Lower panel: Te 5$p_x+p_y$ (dashed line) and $p_z$ (solid line) PDOS.}
\label{DOS}
\end{figure}

In Fig. \ref{DOS}, we show the calculated density of states (DOS); the results from GGA and GGA+SO calculations are shown in the upper and lower panels of Fig. \ref{DOS}(a), respectively. Clearly the entire valence band region is dominated by the Pd 4$d$ partial DOS (PDOS). Te 5$p$ contributions are shown after rescaling the PDOS by 5 times for better comparison. The near Fermi energy states are contributed by hybridized antibonding states. In Fe-based superconductors, the near Fermi level region is essentially contributed by Fe 3$d$ states \cite{Ca122}. However, Pd and Te contributions at the Fermi level in PdTe is comparable. It was found that ligands play an important role in the electronic structure of Fe-based systems \cite{Ca122}. The present results suggest that covalency is stronger in the electronic structure of PdTe. The bonding states appear between 3 - 7 eV binding energies where Te 5$p$ states have large contributions. The energy regime 2 - 4 eV is essentially contributed by the Pd 4$d$ states. Inclusion of spin-orbit coupling (SOC) term has almost negligible effect in the entire energy region except a splitting of the bands between 2 - 4 eV dominated by Pd 4$d$ states; a splitting of the peak around 3 eV.

Due to the trigonal prismatic arrangement of six Te atoms around each Pd atoms, the five-fold degeneracy of Pd 4\textit{d} bands splits into two sets of doubly-degenerate $E_{2g}$ ($d_{x^2-y^2}$, $d_{xy}$) and $E_{1g}$ ($d_{yz}$, $d_{zx}$)) bands and one non-degenerate $A_{1g}$ ($d_{z^2}$) band. Similarly, the three-fold degeneracy of the Te 5\textit{p} bands splits into a doubly-degenerate $E_{1u}$ ($p_x$,$p_y$) and one non-degenerate $A_{2u}$ (= $p_z$) bands \cite{TrigonalPrism}. The calculated GGA+SO results with these symmetries are shown in Fig. \ref{DOS}(b). The DOS distributions of $E_{1g}$ and $E_{2g}$ bands form sharp features dominated in the vicinity of 2.4 eV and 3 eV binding energies, respectively. The $A_{1g}$ contributions are spread over a large energy range down to about 5 eV. Close to the Fermi level, the $E_{g}$ states have almost equal DOS contributions, while $A_{1g}$ contribution is negligible. Te 5\textit{p} contributions are small near Pd 4$d$ $E_{g}$ peaks (energy regime of 2 - 3.5 eV). Te 5$p_z$ states are strongly hybridized with Pd 4$d_{z^2}$ states forming bonding states at around 4 eV. The bonding $p_x + p_y$ states appear at relatively higher binding energies.

\begin{figure}
\includegraphics[width=\linewidth]{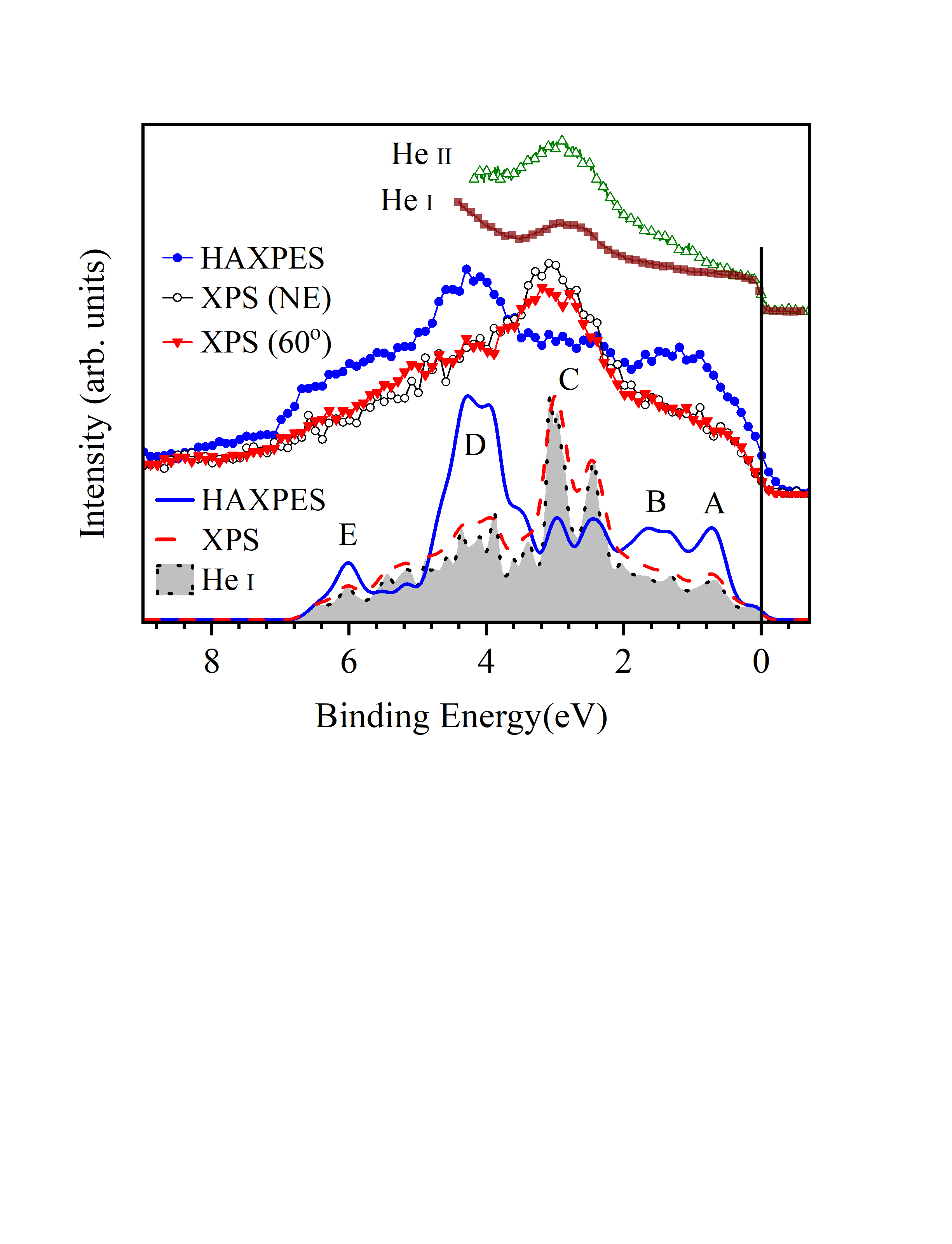}
\vspace{-36ex}
\caption{Experimental valence band spectra collected using 21.2 (He {\scriptsize I}), 40.8 ({\scriptsize II}), 1486.6 (XPS) and 5947 (HAXPES) eV photon energies. Simulation of experimentally measured spectra using atomic photoemission cross-sections are shown at the lower panel.}
\label{VBW}
\end{figure}

In Fig. \ref{VBW}, we show the valence band photoemission spectra collected using different photon energies. Calculated spectral functions using PDOS in Fig. \ref{DOS} and corresponding photoemission cross sections \cite{Yeh} are shown in the lower panel of the figure. Multiple features are observed with significant change in relative intensity of the features with the change in photon energy. Finite spectral weight close to the Fermi level is observed in all the cases indicating metallic ground state of the material as also observed in the calculated results. For better discussion, we have marked the features in the calculated spectra by A ... E in the figure. Due to high matrix cross section\cite{Yeh} for Pd 4$d$ states in comparison to Te 5$p$ state for the spectroscopy with ultraviolet photon energies and significant enhancement of cross section in the He {\scriptsize II} energy compared to He {\scriptsize I} case, the feature, C at 3 eV binding energy is attributed to primarily Pd 4$d$ $E_{1g}$ contributions as also evident in the calculated spectra. In the CXPS data, this peak is the most intense one as expected due to one order of magnitude higher cross-section of Pd 4$d$ states relative to Te 5$p$ cross section. A change in emission angle from normal emission geometry makes the technique surface sensitive which leads to a small reduction in intensity of this feature. This suggests bulk property of this feature. Clearly, the cleaved surface has relatively more Te contribution.

We compare the XPS valence band spectra with the HAXPES data which has dominant bulk contributions. Since the relative cross-section of Te 5$p$ states enhance at 6 keV photon energies compared to XPS case, the enhancement of intensities in the HAXPES data is attributed to such cross-section change in addition to the change in surface sensitivity. As the incident photons in the HAXPES study were polarized and the sample is a high quality single crystal, we considered the polarization induced change in cross-section to simulate the HAXPES spectrum as given in the Ref. \cite{Ca122-Ram}; this is not required for XPS and UVPES studies. To get the simulated spectral functions, we first added the matrix cross section weighted PDOS of Pd 4\textit{d} and Te 5\textit{p} states and then after multiplying the Fermi-Dirac distribution function, the final calculated spectra were convoluted  with Gaussian function with 400 meV FWHM value. The obtained simulation results are shown in the lower panel of the figure exhibiting good representation of the experimental spectra. The calculated HAXPES spectrum represented by solid line provide a good description of the experimental features. The features D and E are the bonding bands having $A_{1g}$ and $p_x + p_y$ symmetries as found in the calculated results. The features, A and B appearing between 0 - 2 eV binding energies are the antibonding states with large covalency.

\begin{figure}
\includegraphics[width=\linewidth]{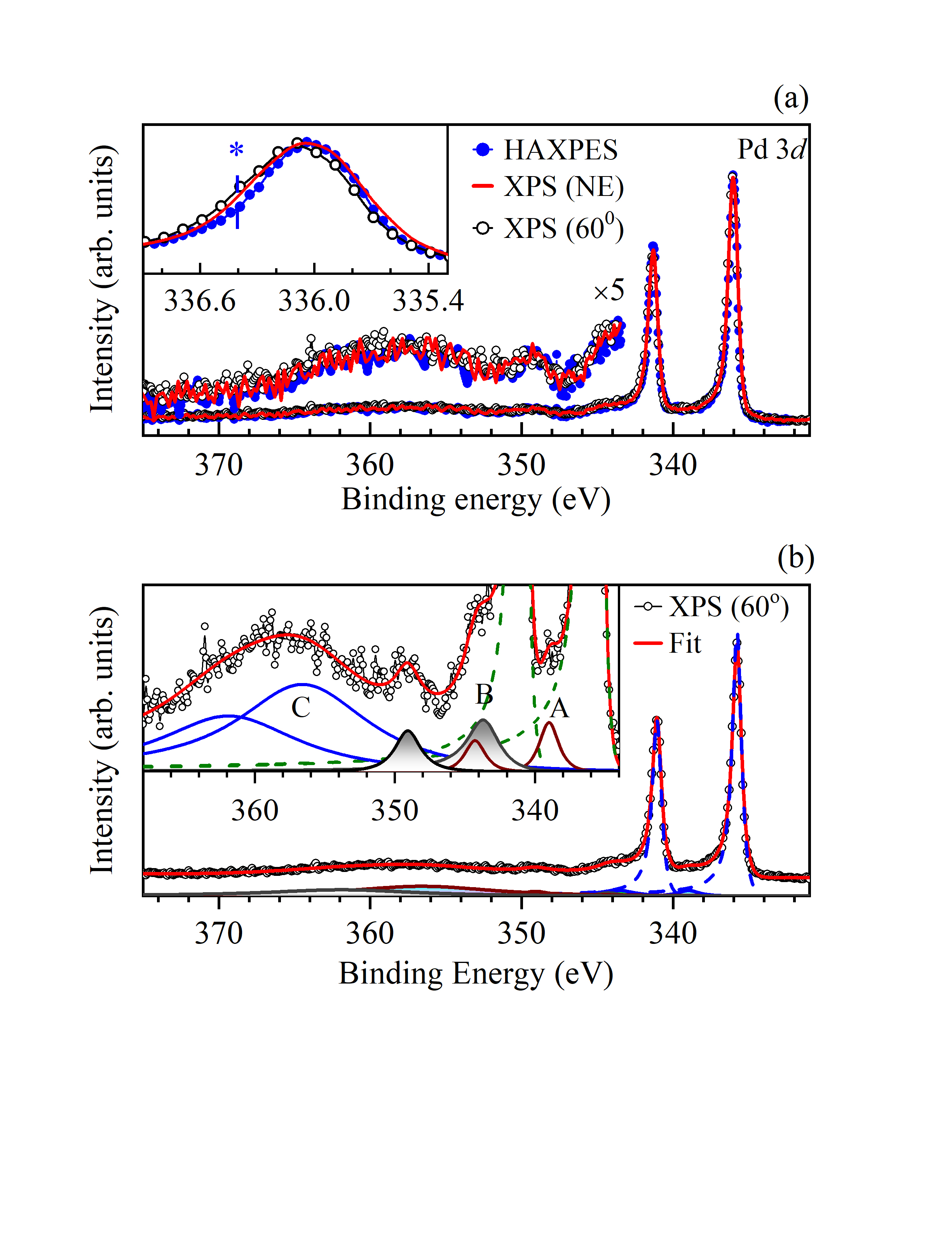}
\vspace{-12ex}
\caption{(a) Pd 3\textit{d} core level spectra collected using hard and soft $x$-ray photoemission spectroscopy. Inset: 3$d_{5/2}$ feature in an expanded energy scale. (b) Simulated Pd 3\textit{d} spectrum. Component peaks are shown in the bottom pane. Inset: data in (b) in an expanded intensity scale.}
\label{Pd3d}
\end{figure}

The surface-bulk differences are studied employing core level spectroscopy. The HAXPES data is collected at normal emission geometry which has the highest bulk sensitivity; escape depth is about 39 \AA for Pd 3$d$ electrons. CXPS spectra were collected at normal emission and 60$\degree$ angled emission geometries having escape depth of about 17 \AA\ and 8.5 \AA, respectively. In Fig. \ref{Pd3d}(a), we show the Pd 3$d$ HAXPES and CXPS spectra exhibiting multiple distinct features. The overall shape of the spectra is almost identical in every case. Two sharp and intense features observed at 336.0 eV and 341.3 eV correspond to Pd 3$d_{5/2}$ and 3$d_{3/2}$ spin-orbit split photoemissions. There is a broad hump at the higher binding energy side of these main peaks. The overall spectral weight distributions are not influenced by the change of surface sensitivity suggesting that the surface-bulk difference for Pd 3$d$ photoemission is not significant. Therefore, multiple features observed in the spectral functions are attributed to different final state peaks. 

In the inset of Fig. \ref{Pd3d}(a), we show the zoomed view of the Pd 3$d_{5/2}$ peak. The HAXPES spectrum exhibits a single peak structure with asymmetry towards higher binding energies as observed in metallic systems due to low energy excitations across the Fermi level. The CXPS spectrum at normal emission exhibits a slightly larger width which may appear due to resolution broadening. Interestingly, the 60$\degree$ emission angled data show enhancement of intensity at the higher binding energy side which manifests as an overall shift of the spectra towards higher binding energies. Considering that the energy resolution and other experimental parameters are identical in both the CXPS cases, the intensities marked by '$\star$' in the inset are attributed to the change in surface sensitivity of the technique and is assigned as surface feature.

In order to identify the constituent peaks in the spectra, we have simulated the experimental data using a set of asymmetric Gaussian-Lorentzian (GL) product functions considered for features distinctly observed in the experimental data. In Fig. \ref{Pd3d}(b), we show the simulated data of Pd 3$d$ 60$\degree$-angled emission CXPS spectrum. The fittings are done using least square error method with a constraint that the width of the spin-orbit split pair is same and ratio of the integrated area under the curve follows the degeneracies of the features. The intense broad feature, C at 357 eV and 362 eV are attributed to plasmon excitations. The features within the binding energy range 336-350 eV are contributed by two pairs of the final state peaks associated to the spin-orbit spilt photoemission signals. The features, A and B are associated to 3$d_{5/2}$ photo-excitations and appear at 339 and 343.8 eV; signature of satellites are also observed in earlier studies of Pd-based florides \cite{Pd-SAT}. These satellites appear due to the final state effect in the presence of finite electron correlation \cite{satellites}. Evidently, electron correlation cannot be neglected to derive the electronic properties of these materials which is in line with the unconventional nature of the superconductivity observed in this system \cite{PdTe-SREP-Chapai}.

\begin{figure}
\includegraphics[width=\linewidth]{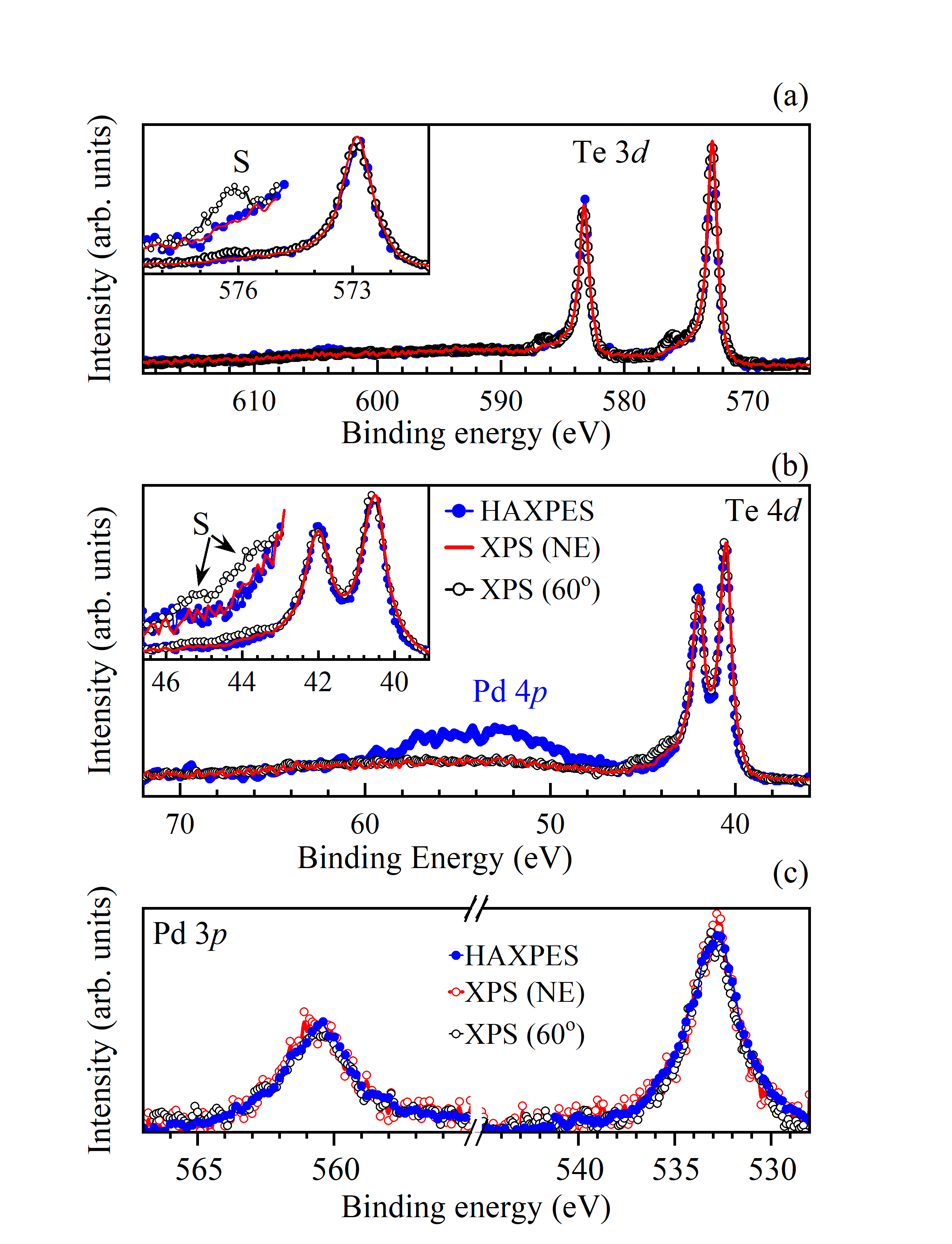}
\caption{(a) Te 3$d$ spectra at different photon energies. Inset: 3$d_{5/2}$ signal in an expanded energy scale. (b) Te 4$d$ spectra at different photon energies. Inset: 4$d_{5/2}$ and 4$d_{3/2}$ main peak region in an expanded energy scale. (c) Pd 3$p$ spectra at different photon energies.}
\label{Te_CXPS}
\end{figure}

In Fig. \ref{Te_CXPS}(a), we study the Te 3$d$ HAXPES and CXPS core level spectra measured at different experimental geometries. Normal emission HAXPES and CXPS spectra show two sharp and intense features at around 572.8 and 583.3 eV along with intensities in the broad energy range, 590-610 eV of very similar spectral shape. Interestingly, the 60$\degree$-angled emission CXPS spectrum exhibits distinct additional features at 576.1 and 586.5 eV associated to Te 3$d_{5/2}$ and 3$d_{3/2}$ photoemissions, respectively; expanded view of the 3$d_{5/2}$ case is shown in the inset of Fig. \ref{Te_CXPS}(a). The presence of these peaks in the angled emission CXPS case suggests surface nature of these signals. The appearance at such a higher binding energy is puzzling as Te is more electronegative than Pd in this material. Such an energy shift to higher binding energy suggests relatively more positive valency at the surface compared to the bulk.

In Fig. \ref{Te_CXPS}(b), we show the Te 4$d$ core level HAXPES and CXPS spectra. Two sharp and intense features within 40-44 eV range are identified as Te 4$d$ spin-orbit split features; Te 4$d_{5/2}$ and 4$d_{3/2}$ peaks. The broad features within the energy window 48-60 eV is identified as Pd 4$p$ peaks. Matrix cross section of the Pd 4$p$ photoexcitation at Al $K\alpha$ $x$-ray energy is very small and hence the Pd 4$p$ intensity is weak in the CXPS data in comparison to the HAXPES spectrum. In addition, distinct signatures of the surface features are observed in the experimental data. These results suggest that the surface electronic structure may be dominated by the Te-contributions as also seen in the valence band spectra. In the inset of the Fig. \ref{Te_CXPS}(b), we show the zoomed view of the Te 4$d$ spectral functions. Within 43-46 eV range, a clear increase of spectral weight is observed in the most surface sensitive case. Binding energy difference ($\sim$ 3 eV) of these features from the main peak is similar to the values observed in the Te 3$d$ case.

To confirm that the energy position of the above discussed surface features are not linked to the undetected trace of adsorbed oxygen at the cleaved surface, we investigate the Pd 3\textit{p} CXPS and HAXPES spectra in Fig. \ref{Te_CXPS}(c) as Pd photoemission signals also will be influenced by the presence of surface oxygens. The experimental data shown in Fig. \ref{Te_CXPS}(c) does not show any additional peak; all the spectra look almost identical. This is very similar to the scenario observed for Pd 3$d$ photoemission shown in Fig. \ref{Pd3d}. Clearly, the role of adsorbed oxygen, if there is any, may not be significant and the surface feature observed in Te core level spectra has different origin.

\begin{figure}
\includegraphics[width=\linewidth]{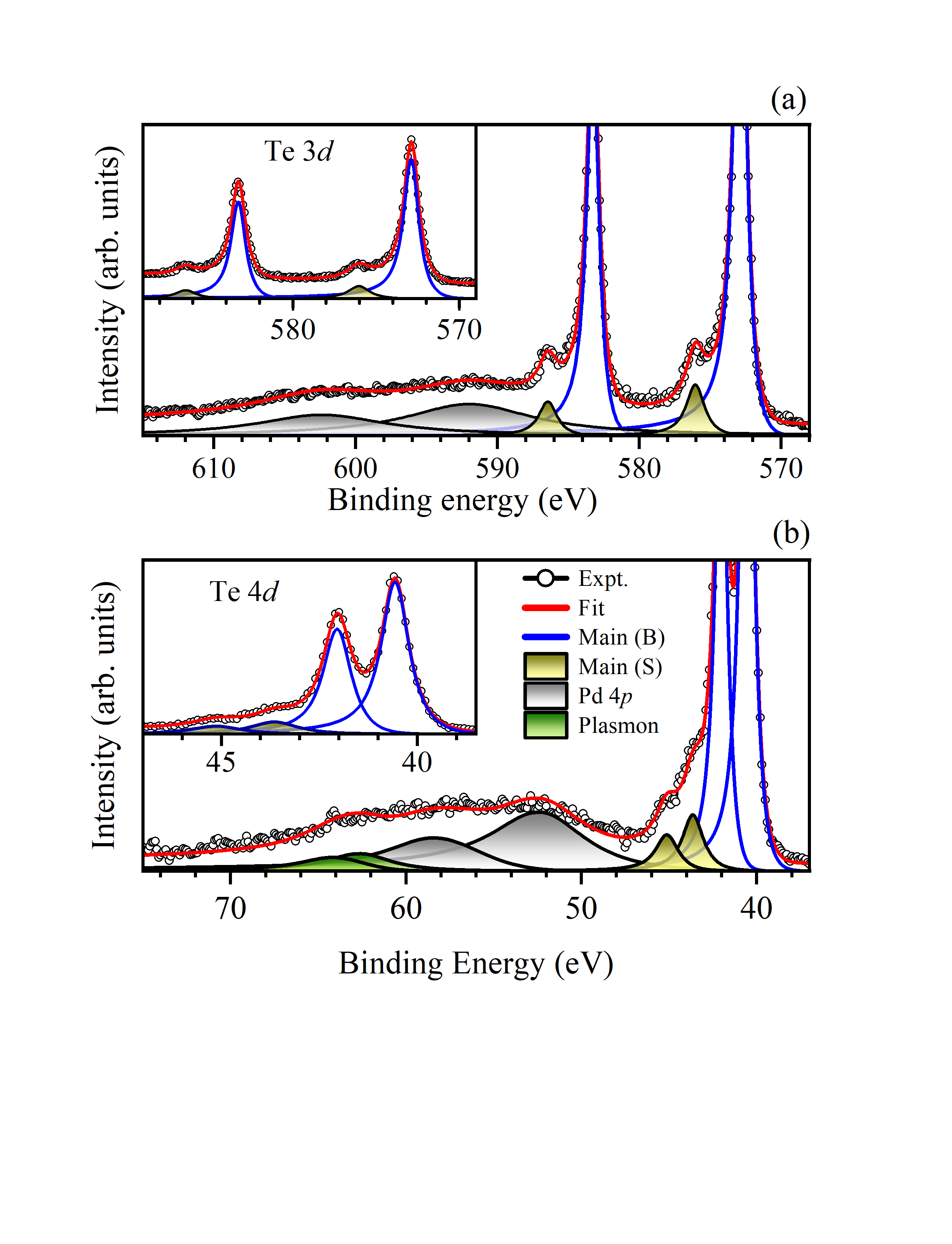}
\vspace{-12ex}
\caption{(a) Simulated Te 3$d$ spectrum measured at 60$\degree$-angled emission geometry. The component peaks are shown in the bottom panel. Inset: Te 3$d$ main peak region in an expanded scale. (b) Simulated Te 4$d$ spectrum collected at 60$\degree$-angle. Inset: Main peak region in an expanded scale.}
\label{Te_Sim}
\end{figure}

In order to find the features constituting the Te core level spectra shown in Fig. \ref{Te_CXPS}, simulation of the surface sensitive case were carried out following the procedure used for Pd core level. The results are shown in Fig. \ref{Te_Sim}. Three pairs of peaks were required to simulate the Te 3$d$ spectrum. The binding energy of plasmon features are 592.0 and 602.0 eV. The binding energies for the spin-orbit split surface features are 575.8 and 586.2 eV, which are 3 eV away from the bulk signals. The binding energy of the bulk core level peaks is close to those observed in Te metal suggesting weak negative valency of these atoms in this highly covalent material. The binding energy of the surface features shows significantly higher valency of the surface Te atoms. Similar width of the surface and bulk peaks suggests that the effect of surface disorder, if there is any, is not significant. Simulation results for the Te 4$d$ core level spectrum also exhibit similar scenario. The binding energy of spin-orbit split Pd 4$p$ features are observed at 53.0 and 59.0 eV. Plasmon features corresponding to Te 4$d$ photoemission are observed in the energy range 60 - 70 eV. While the bulk peaks appear at about 40.6 and 42.1 eV for the spin-orbit split Te 4$d$ features, the surface peaks related to the corresponding photoemission are observed at 43.7 and 45.2 eV; about 3 eV away from the main peak as also found in Te 3$d$ case.

Core level shift of the surface atoms with respect to bulk are usually captured by two models: (a) the charge transfer model \cite{Ca122-Ram,Ba122,Ca122}, where the translational symmetry breaking at the surface leads to renormalization of covalency of the surface atoms thereby a change in effective Madelung potential due to transfer of charges between the constituent atoms. The energy shift of the core levels will be opposite for atoms with different ionicity. (b) The other model is the narrowing of the valence band \cite{Hufner,Surface-BiPd,Zhang-Be0001}, where the narrowing of the bands shifts the Fermi level and hence all the core levels will shift in one direction by the similar amount in absence of charge transfer. In the present case, we observe a weak shift of Pd towards higher binding energy. Curiously, the Te peaks shifts significantly towards higher binding energy although it is more electronegative than Pd. Clearly a charge transfer model is not applicable in the present case. Highly anomalous energy shift of Pd and Te peaks rules out the second possibility too. It is to note here that varying band narrowing effect of the surface and sub-surface layers can lead to an unequal shift \cite{Hufner}. However, such scenario can also be ruled out in the present case, as the band narrowing of more than half filled Te 5$p$ valence orbitals is expected to shift core levels towards the Fermi level in contrast to the experimentally observed scenario.

Energy of the surface states is sensitive to the arrangement of surface atoms. One reason for such anomalous spectral modification at the surface may be the presence of vacancies, which may lead to a significant enhancement of effective Madelung potential of the surface Te atoms. Vacancies often lead to interesting unusual properties in a material \cite{CaB6}. Moreover, cleaved surfaces are expected to have half of the atoms distributed in the two surfaces it generates \cite{Hoffman}. Therefore, breaking of translational symmetry, which usually leads to unsaturated bonds and the absence of half of the surface layer will reduce energy by rearranging themselves, known as surface reconstruction. Such reconstruction usually results into a shift of the surface core levels where the magnitude and sign of the surface peak shift is highly sensitive to the rearrangement of surface atoms \cite{Zhang-AuSi}. Clearly more studies are required to unravel the underlying physics.

\section{Conclusion}

In summary, we studied the detailed electronic structure of PdTe using depth-resolved photoemission spectroscopy and density functional theory. Large density of states with flat intensities in a wide energy range in the vicinity of the Fermi level suggests highly metallic character of the valence states. The states at the Fermi level possess highly mixed character suggesting importance of covalency in the exotic electronic properties of this material. Core level spectra show multiple features and signature of plasmon excitations. Pd core level spectra exhibit satellite features due to final state effects indicating importance of electron correlation in the electronic structure. Evidently, the exoticity in this material involves competing interactions of covalency and electron correlation. The depth resolved spectra reveals unusual surface-bulk differences in the electronic structure; the surface electronic structure is dominated by the Te-contributions. The surface features found in the core level spectra are anomalous and may be attributed to the surface reconstruction and/or vacancies at the surface.

\section{Acknowledgement}

The authors acknowledge the financial support from the Department of Science and Technology, Govt. of India under India-DESY program and Department of Atomic Energy, Govt. of India.


\begin{thebibliography}{99}

\bibitem{TISC-1}
L. A. Wray, S. Y. Xu, Y. Xia, Y. S. Hor, D. Qian, A. V. Fedorov, H. Lin, A. Bansil, R. J. Cava, and M. Z. Hasan, Nat. Phys. \href{https://doi.org/10.1038/NPHYS1762}{\textbf{6}, 855 (2010)}.

%%
\bibitem{TISC-2}
Shruti, V. K. Maurya, P. Neha, P. Srivastava, and S. Patnaik, Phys. Rev. B \href{https://doi.org/10.1103/PhysRevB.92.020506}{\textbf{92}, 020506(R) (2015)};
%
C. Q. Han, H. Li, W. J. Chen, F. Zhu, M. Y. Yao, Z. J. Li, M. Wang, B. F. Gao, D. D. Guan, C. Liu, C. L. Gao, D. Qian, and J. F. Jia, Appl. Phys. Lett. \href{https://doi.org/10.1063/1.4934590}{\textbf{107}, 171602 (2015)}.

%%
\bibitem{TISC-3}
A. S. Erickson, J.-H. Chu, M. F. Toney, T. H. Geballe, and I. R. Fisher, Phys. Rev. B \href{https://doi.org/10.1103/PhysRevB.79.024520}{\textbf{79}, 024520 (2009)};
%
A. Maiti, R. P. Pandeya, B. Singh, K. K. Iyer, A. Thamizhavel, and K. Maiti Phys. Rev. B \href{https://doi.org/10.1103/PhysRevB.104.195403}{\textbf{104}, 195403 (2021)};
%
A. Maiti, A, Singh; K. K. Iyer, and A. Thamizhavel, Appl. Phys. Lett.
\href{https://doi.org/10.1063/5.0086644}{\textbf{120}, 112102 (2022)}

%%
\bibitem{TISC-4}
B. Joshi, A. Thamizhavel, and S. Ramakrishnan, Phys. Rev. B \href{https://doi.org/10.1103/PhysRevB.84.064518}{\textbf{84}, 064518 (2011)};
%
A. Pramanik, R. P. Pandeya, D. V. Vyalikh, A. Generalov, P. Moras, A. K. Kundu, P. M. Sheverdyaeva, C. Carbone, B. Joshi, A. Thamizhavel, S. Ramakrishnan, and K. Maiti, Phys. Rev. B \href{https://doi.org/10.1103/PhysRevB.103.155401}{\textbf{103}, 155401 (2021)};
%


\bibitem{PdTe-SREP-Chapai}
Ramakanta Chapai, P. V. Sreenivasa Reddy, Lingyi Xing, David E. Graf, Amar B. Karki, Tay‑Rong Chang, and Rongying Jin, Sci. Rep. \href{https://doi.org/10.1038/s41598-023-33237-5}{\textbf{13}, 6824 (2023)}.

\bibitem{PdTe-PRL-Hasan}
X. P. Yang, Y. Zhong, S. Mardanya, T. A. Cochran, R. Chapai, A. Mine, J. Zhang, J. S\'{a}nchez-Barriga, Zi-Jia Cheng, O. J. Clark, Jia-Xin Yin, J. Blawat, G. Cheng, I. Belopolski, T. Nagashima, S. Najafzadeh, S. Gao, Nan Yao, A. Bansil, R. Jin, T.-R. Chang, S. Shin, K. Okazaki, and M. Z. Hasan, Phys. Rev. Lett. \href{https://doi.org/10.1103/PhysRevLett.130.046402}{\textbf{130}, 046402 (2023)}.

\bibitem{Surface-Reconstruction}
Charles B. Duke, Chem. Rev. \href{https://pubs.acs.org/doi/pdf/10.1021/cr950212s}{\textbf{96}, 1237-1259 (1996)}.

\bibitem{TI-Surface}
M. Z. Hasan and C. L. Kane, Rev. Mod. Phys. \href{https://journals.aps.org/rmp/pdf/10.1103/RevModPhys.82.3045}{\textbf{82}, 3045-3067 (2010)};
%
X.-L. Qi and S.-C. Zhang, Rev. Mod. Phys. \href{https://journals.aps.org/rmp/pdf/10.1103/RevModPhys.83.1057}{\textbf{83}, 1057 (2011)}.

\bibitem{Surface-Applications}
G. A. Somorjai, and Y. Li, PNAS \href{https://www.pnas.org/content/pnas/108/3/917.full.pdf}{\textbf{108} (3) 917-924 (2011)}.

\bibitem{Surface}
Michael Horn-von Hoegen, Zeitschrift für Kristallographie \href{https://www.uni-due.de/imperia/md/content/ag-hvh/vorlesung/hvh_zkrist_214_591_1999.pdf}{\textbf{214},  591-629 and 684-721 (1999)};
%
{\O}. Fischer,  M. Kuglar, I. Maggio-Aprile, and Chrirsophe Berthod, Rev. Mod. Phys. \href{https://journals.aps.org/rmp/pdf/10.1103/RevModPhys.79.353}{\textbf{79}, 353 (2007)};
%
Jennifer E Hoffman, Rep. Prog. Phys. \href{https://iopscience.iop.org/article/10.1088/0034-4885/74/12/124513/pdf}{\textbf{74}, 124513 (2011)}.

\bibitem{Surface-PES}
K. Maiti, P. Mahadevan, and D. D. Sarma, Phys. Rev. Lett. \href{https://doi.org/10.1103/PhysRevLett.80.2885}{\textbf{80}, 2885 (1998)};
%
F. Reinert, and S. Hufner, New J. Phys. \href{https://iopscience.iop.org/article/10.1088/1367-2630/7/1/097/pdf}{\textbf{7}, 97 (2005)};
%
D. Biswas, S. Thakur, K. Ali, G. Balakrishnan, and K. Maiti, Sci. Rep. \href{https://www.nature.com/articles/srep17351.pdf}{\textbf{5}, 10260 (2015)}.

\bibitem{Hufner}
S. H$\ddot{u}$fner, \textit{Photoelectron Spectroscopy} (\textit{Springer-Verlag, Berlin}, \href{https://www.springer.com/gp/book/9783540418023}{1995}).


\bibitem{Karki-PdTe}
A. B. Karki, D. A. Browne, S. Stadler, J. Li, and R. Jin, J. Phys. Condens. Matter \href{https://iopscience.iop.org/article/10.1088/0953-8984/24/5/055701/pdf}{{\bf 24}, 055701 (6pp) (2012)}.

\bibitem{Brajesh-PdTe}
B. Tiwari, R. Goyal, R. Jha, A. Dixit, and VPS Awana, Supercond. Sci. Technol. \href{https://iopscience.iop.org/article/10.1088/0953-2048/28/5/055008/pdf}{{\bf 28}, 055008 (5pp) (2015)}.

\bibitem{Karki-FePdTe}
A. B. Karki, V. O. Garlea, R. Custelcean, S. Stadler, W. W. Plummer, and R. Jin, PNAS \href{https://www.pnas.org/content/pnas/110/23/9283.full.pdf}{{\bf 23}, 110, 9283-9288 (2013)}.

\bibitem{Jha-NiPdTe}
R. Jha, R. Goyal, B. Tiwar, and V. P. S. Awana, Supercond. Sci. Technology \href{https://iopscience.iop.org/article/10.1088/0953-2048/29/7/075008/pdf}{{\bf 29}, 075008 (9pp) (2016)}.

\bibitem{PdTe-Ram}
R. P. Pandeya, Arindam Pramanik, Pramita Mishra, A. Thamizhavel, and Kalobaran Maiti, AIP Conf. Proc. \href{https://aip.scitation.org/doi/pdf/10.1063/1.5113178}{{\bf 2115}, 030339  (2019)}.

\bibitem{Matthias-PdTe}
B. Matthias, Phys. Rev. \href{https://journals.aps.org/pr/abstract/10.1103/PhysRev.90.487}{\textbf{90}, 487 (1953)}; B. Matthias, Phys. Rev. \href{https://journals.aps.org/pr/pdf/10.1103/PhysRev.92.874}{\textbf{92}, 874 (1953)}.

\bibitem{HHosono2015}
H. Hosono, K. Tanabe, E. T.-Muromachi, H. Kageyama, S. Yamanaka, H. Kumakura, M. Nohara, H. Hiramatsu, and S. Fujitsu, Sci. Technol. Adv. Mater. \href{https://iopscience.iop.org/article/10.1088/1468-6996/16/3/033503/pdf}{\textbf{16}, 033503 (2015)};
%
K. Maiti, Pramana, \href{https://doi.org/10.1007/s12043-015-0992-x}{\textbf{84}, 947 (2015)}.

\bibitem{Ekuma-Wannier-PdTe}
C. E. Ekuma, C.-H. Lin, J. Moreno, W. Ku, and M. Jarrell, J. Phys.: Condens. Matter \href{https://iopscience.iop.org/article/10.1088/0953-8984/25/40/405601/pdf}{\textbf{25}, 405601 (6pp) (2013)}.

\bibitem{Cao-DFT-PdTe}
J.-J. Cao, and X. F. Gou, Physica C: Superconductivity and its applications \href{https://www.sciencedirect.com/science/article/pii/S0921453415003329}{\textbf{520}, 19–23 (2016)}.


\bibitem{Wien2k}
P. Blaha, K. Schwarz , G. Madsen, D. Kvasnicka and J. Luitz, \textit{WIEN2k: An Augumented Plane Wave + Local Orbitals Program  For Calculating Crystal Properties (Karlheinz Schwarz, Techn. Universit\"{a}t Wien, Austria)} \href{https://wiki.cse.ucdavis.edu/_media/support:hpc:software:wien2k_usersguide.pdf}{\textbf{ ISBN 3-9501031-1-2} (2001)}.

\bibitem{PBE_GGA}
J. P. Perdew, K. Burke and M. Ernzerhof, Phys. Rev. Lett. \href{https://journals.aps.org/prl/pdf/10.1103/PhysRevLett.77.3865}{\textbf{77}, 1396 (1997)}.

\bibitem{Ca122}
G. Adhikary, N. Sahadev, D. Biswas, R. Bindu, N. Kumar, A. Thamizhavel, S. K. Dhar, and K. Maiti, J. Phys.: Condens. Matter, \href{https://doi.org/10.1088/0953-8984/25/22/225701}{\textbf{25}, 225701 (2013)};
%
G. Adhikary, D. Biswas, N. Sahadev, S. Ram, V. Kanchana, C. S. Yadav, P. L. Paulose, and K. Maiti, J. Appl. Phys. \href{https://doi.org/10.1063/1.4827192}{\textbf{114}, 163906 (2013)};
%
G. Adhikary, D. Biswas, N. Sahadev, R. Bindu, N. Kumar, S. K. Dhar, A. Thamizhavel, and K. Maiti, J. Appl. Phys. \href{https://aip.scitation.org/doi/pdf/10.1063/1.4869397}{\textbf{115}, 123901 (2014)};
%
K. Ali and K. Maiti, Sci. Rep. \href{https://doi.org/10.1038/s41598-017-06591-4}{\textbf{7}, 6298 (2018)}.

\bibitem{TrigonalPrism}
R. Huisman, R. de Jonge, C. Haas, and F. Jellinek, J. Solid State Chemistry
\href{https://www.sciencedirect.com/science/article/pii/0022459671900077}{\textbf{3}, 56-66 (1971)}.

\bibitem{Yeh}
J. J. Yeh and I. Lindau, \href{https://www.sciencedirect.com/science/article/pii/0092640X85900166?via\%3Dihub}{\textbf{32}, 1-155 (1985)}.

\bibitem{Ca122-Ram}
R. P. Pandeya, A. Pramanik, A. P. Sakhya, A. Thamizhavel, and K. Maiti, J. Phys: Condens. Matter (Letter) \href{https://iopscience.iop.org/article/10.1088/1361-648X/ab86f0/pdf}{{\bf 32}, 33LT01 (6pp) (2020)}.

\bibitem{Pd-SAT}
A. Tressad, S. Khairou, H. Touha, and N. Watana, Z. Anorg. Allg. Chem. \href{https://onlinelibrary.wiley.com/doi/epdf/10.1002/zaac.19865400932}{\textbf{540/541}, 291-299 (1986)}.

\bibitem{satellites}
M. Imada, A. Fujimori, and Y. Tokura, Rev. Mod. Phys. \href{https://doi.org/10.1103/RevModPhys.70.1039}{\textbf{70}, 1039 (1998)};
K. Maiti and R. S. Singh, Phys. Rev. B \href{https://doi.org/10.1103/PhysRevB.71.161102}{\textbf{71}, 161102(R) (2005)};
K. Maiti and R. S. singh, Sci. Lett. J. \href{http://www.cognizure.com/abs/200638433.aspx}{\textbf{3}, 55 (2014)};
K. Maiti, P. Mahadevan, and D. D. Sarma, Phys. Rev. B \href{https://doi.org/10.1103/PhysRevB.59.12457}{\textbf{59}, 12457 (1999)}.

\bibitem{Ba122}
S. de Jong, Y. Huang, R. Huisman, F. Massee, S. Thirupathaiah, M. Gorgoi, F. Schaefers, R. Follath, J. B. Goedkoop, and M. S. Golden, Phys. Rev. B \href{https://iop.fnwi.uva.nl/cmp/docs/Golden/DeJong_PRB79(2009)115125.pdf}{\textbf{79}, 115125 (2009)}.


%\bibitem{Ca122-Ram-ARPES}
%R. P. Pandeya, A. P. Sakhya, S. Datta, T. Saha, G de Ninno, R. Mondal, A. Thamizhavel, and K. Maiti %(unpublished).

\bibitem{Surface-BiPd}
A. Pramanik, R. P. Pandeya, K. Ali, B. Joshi, I. Sarkar, P. Moras, P. M. Sheverdyaeva, A. K. Kundu, C. Carbone, A. Thamizhavel, S. Ramakrishnan, and K. Maiti, Phys. Rev. B \href{https://journals.aps.org/prb/pdf/10.1103/PhysRevB.101.035426}{\textbf{101}, 035426 (2020)}.

\bibitem{Zhang-Be0001}
 L. I. Johansson, P.-A. Glans, and T. Balasubramanian, Phys. Rev. B \href{https://journals.aps.org/prb/pdf/10.1103/PhysRevB.58.3621}{\textbf{58}, 3621 (1998)}.

\bibitem{CaB6}
K. Maiti, Europhysics Letters, \href{https://doi.org/10.1209/0295-5075/82/67006}{\textbf{82}, 67006 (2008)};
%	
K. Maiti, V. R. R. Medicherla, S. Patil and R. S. Singh, Phys. Rev. Letts. \href{https://doi.org/10.1103/PhysRevLett.99.266401}{\textbf{99}, 266401 (2007)}.

\bibitem{Hoffman}
Jennifer E. Hoffman, Rep. Prog. Phys. \href{https://doi.org/10.1088/0034-4885/74/12/124513}{\textbf{74}, 124513 (2011)}.

\bibitem{Zhang-AuSi}
H. M. Zhang, T. Balasubramanian, and R. I. G. Uhrberg, Phys. Rev. B \href{https://journals.aps.org/prb/pdf/10.1103/PhysRevB.65.035314}{\textbf{65}, 035314 (2001)}.

\end{thebibliography}
\end{document}